\documentclass[twocolumn,showkeys,showpacs,preprintnumbers,prd,superscriptaddress,nofootinbib]{revtex4-1}
\usepackage{verbatim}
\usepackage[T1]{fontenc}
\usepackage[utf8]{inputenc}
\usepackage[american]{babel}
\usepackage{epsfig}
\usepackage{booktabs}
\usepackage{multirow}
\usepackage{dcolumn}
\usepackage{amsmath}
\usepackage{mathtools}
\usepackage{amsfonts}
\usepackage{amssymb}
\usepackage{ulem}
\usepackage{epstopdf}
\usepackage{bm}
\usepackage{siunitx}
\usepackage{braket}
\usepackage{enumitem}
\usepackage{soul}
\usepackage[table]{xcolor}
\usepackage{color}
\usepackage{transparent}
\usepackage{pifont}
\usepackage{enumitem}

\definecolor{darkblue}{rgb}{0.0, 0.0, 0.55}
\definecolor{darkred}{rgb}{0.55, 0.0, 0.0}

\usepackage{hyperref}
\hypersetup{
    colorlinks=true, 
    linkcolor=darkblue,
    citecolor=darkblue,
    urlcolor=darkblue}

\begin{document}

\title{Non-Linear Matter Power Spectrum Modeling in Interacting Dark Energy Cosmologies}

\author{Emanuelly Silva}
\email{emanuelly.santos@ufrgs.br}
\affiliation{Instituto de F\'{i}sica, Universidade Federal do Rio Grande do Sul, 91501-970 Porto Alegre RS, Brazil}

\author{Ubaldo Z\'{u}ñiga-Bolaño}
\email{ubaldo.zuniga@udea.edu.com}
\affiliation{Instituto de F\'{i}sica, Universidad de Antioquia, Cl. 67 \#53-108, Medell\'{i}n, Colombia}

\author{Rafael C. Nunes}
\email{rafadcnunes@gmail.com}
\affiliation{Instituto de F\'{i}sica, Universidade Federal do Rio Grande do Sul, 91501-970 Porto Alegre RS, Brazil}
\affiliation{Divis\~{a}o de Astrof\'{i}sica, Instituto Nacional de Pesquisas Espaciais, Avenida dos Astronautas 1758, S\~{a}o Jos\'{e} dos Campos, 12227-010, S\~{a}o Paulo, Brazil}

\author{Eleonora Di Valentino}
\email{e.divalentino@sheffield.ac.uk}
\affiliation{School of Mathematics and Statistics, University of Sheffield, Hounsfield Road, Sheffield S3 7RH, United Kingdom}

\begin{abstract}
Understanding the behavior of the matter power spectrum on non-linear scales beyond the $\Lambda$CDM model is crucial for accurately predicting the large-scale structure (LSS) of the Universe in non-standard cosmologies. In this work, we present an analysis of the non-linear matter power spectrum within the framework of interacting dark energy-dark matter cosmologies (IDE). We employ N-body simulations and theoretical models to investigate the impact of IDE on these non-linear scales. Beginning with N-body simulations characterized by a fixed parameter space delineated by prior observational research, we adeptly fit the simulated spectra with a simple parametric function, achieving accuracy within 5\%. Subsequently, we refine a modified halo model tailored to the IDE cosmology, exhibiting exceptional precision in fitting the simulations down to scales of approximately 1 h/Mpc. To assess the model's robustness, we conduct a forecast analysis for the Euclid survey, employing our refined model. 
We find that the coupling parameter $\xi$ will be constrained to $\sigma(\xi) = 0.0110$. This marks a significant improvement by an order of magnitude compared to any other current observational tests documented in the literature. These primary findings pave the way for a novel preliminary approach, enabling the utilization of IDE models for observational constraints concerning LSS data on non-linear scales.
\end{abstract}



\maketitle

\section{Introduction}
\label{sec:introduction}

Dark energy (DE) is one of the most enigmatic and perplexing phenomena in the field of astrophysics and cosmology. It may represent a new field that permeates the universe, driving its current accelerated expansion~\cite{DE_obs_1, DE_obs_2}. DE as a cosmological constant ($\Lambda$) represents our simplest understanding of the universe's expansion and evolution in the last two decades. The DE existence and influence on the cosmos are supported by a wealth of observational evidence, Supernova Surveys, Cosmic Microwave Background (CMB), Large-Scale Structure (LSS), Gravitational Lensing, among others~\cite{huterer2017dark, mortonson2013dark}. The success of the $\Lambda$ term in explaining these observations has led to the $\Lambda$CDM scenario being considered the standard model of cosmology.

However, recently, a few tensions and anomalies have emerged as statistically significant when analyzing different datasets. The most noteworthy disagreement lies in the value of the Hubble constant, $H_0$, which exists between the Planck-CMB estimate~\cite{Planck:2018vyg}, assuming the standard $\Lambda$CDM model, and the direct local distance ladder measurements conducted by the SH0ES team~\cite{Riess:2021jrx, Riess_2022, murakami2023leveraging}, reaching a significance of more than 5$\sigma$. Furthermore, within the $\Lambda$CDM framework, the CMB measurements from Planck and ACT-DR6~\cite{Planck:2018vyg, qu2023atacama} provide values of $S_8 = \sigma_8 \sqrt{\Omega_m/0.3}$ in 2-3$\sigma$ statistical tension with those inferred from various weak lensing, galaxy clustering, cluster counts, and redshift-space distortion measurements~\cite{DiValentino:2020vvd, Nunes_2021, DES:2021bvc, Asgari_2021}.
Various other anomalies and tensions have emerged within the $\Lambda$CDM framework in recent years~\cite{Perivolaropoulos_2022,Abdalla:2022yfr}. Motivated by such discrepancies, which are unlikely to disappear completely by introducing multiple systematic errors, there has been widespread discussion in the literature regarding whether new physics beyond the standard cosmological model can resolve these tensions~\cite{DiValentino:2021izs,Knox:2019rjx,Jedamzik:2020zmd,Kamionkowski:2022pkx,Khalife:2023qbu}. Interacting Dark Energy - Dark Matter (IDE) models have been extensively investigated as potential solutions to the current cosmological tensions~\cite{Kumar:2016zpg, Murgia:2016ccp, Kumar:2017dnp, DiValentino:2017iww, Yang:2018ubt, Yang:2018euj, Yang:2019uzo, Kumar:2019wfs, Pan:2019gop, Pan:2019jqh, DiValentino:2019ffd, DiValentino:2019jae, DiValentino:2020leo, DiValentino:2020kpf, Gomez-Valent:2020mqn, Yang:2019uog, Lucca:2020zjb, Martinelli:2019dau, Yang:2020uga, Yao:2020hkw, Pan:2020bur, DiValentino:2020vnx, Yao:2020pji, Amirhashchi:2020qep, Yang:2021hxg, Gao:2021xnk, Lucca:2021dxo, deCruzPerez:2023wzd,Kumar:2021eev,Yang:2021oxc,Lucca:2021eqy,Halder:2021jiv,Bernui_2023, Poulin_2023, lucca2021dark, Zhai_2023, Gariazzo_2022, Di_Valentino_2020, mawas2021interacting, Yang_2021, Lucca_2020, Pan_2020, von_Marttens_2020, vanderWesthuizen:2023hcl,Goh:2022gxo,Gomez-Valent:2022bku,Escamilla:2023shf, Forconi:2023hsj, hoerning2023constraints, Wang:2024vmw,Benisty:2024lmj}.

At the intersection of various cosmic phenomena, especially at the LSS observation level, lies the matter power spectrum, an essential cosmological tool for unraveling the interplay between cosmological parameters and the emergence of LSS. The power spectrum provides a unique probe through which we can peer into the universe’s past and present, revealing the traces of primordial fluctuations, the gravitational dynamics of dark matter (DM), and the elusive nature of DE. On the other hand, is well known that the non-linear power spectrum plays a pivotal role in current LSS cosmological surveys, offering crucial insights into the complex and dynamic behavior of cosmic structures. The non-linear power spectrum is a crucial component in any current LSS cosmological survey because it serves as a bridge between theoretical predictions and observational data. 

The non-linear power spectrum is well modeled for the $\Lambda$CDM model and some of its basic extensions~\cite{Takahashi_2012, matsubara2008non-linear, hand2017extending, vlah2016gaussian, chen2021redshift, baumann2012cosmological, cabass2023snowmass}. However, modeling non-linear regimes in non-standard cosmologies poses a significant challenge. Various methodologies have been employed to address this challenge, including the Effective Field Theory of LSS~\cite{d2021limits, d2021hubble, ivanov2020cosmological, colas2020efficient, d2020limits, bose2018towards}, the Halo models~\cite{bose2023fast, carrilho2022road, bose2020road, gupta2023improved, carrion2024dark, Spurio_Mancini_2022}, and emulators trained via N-body simulation data~\cite{winther2019emulators, fiorini2023fast, mauland2023sesame, saez2024mantis, ruan2024emulator}. These approaches offer promising avenues for improving our understanding of non-linear regimes in cosmology, but further research is needed to explore their capabilities and limitations.

In the context of IDE models, some significant strides have been made in constraining model parameters using the galaxy power spectrum. A notable study by ~\cite{Nunes_2022_IDE} provided an in-depth analysis of the coupling parameter $\xi$, which represents the interaction strength between DE and DM. By combining the full power spectrum from the BOSS DR12 galaxy sample with Planck CMB data, the authors established a lower bound of $\xi > -0.12$ at the 95\% CL. Despite this progress, there remains a paucity of studies that employ non-linear modeling within the IDE framework using real observational data, highlighting a critical area for future research.

Given the imperative to anticipate model behavior on non-linear scales, it becomes essential to employ phenomenological models for the interpretation of current and forthcoming data derived from galaxy clustering and cosmic shear. Exploring this approach is particularly important as it helps bridge the gap between theoretical predictions and observational data. By doing so, we can enhance our understanding of the perturbation dynamics within these complex systems and improve our comprehension of the observational data. In this work, our main aim is to build a robust non-linear $P(k)$ model in the presence of a DE-DM interaction. To this end, we carry out a series of N-body gravity simulations and then we develop these perspectives in two ways. First, we capture the ratio $R(k)=P_{\rm NL}(k)/P_{\rm linear}(k)$ up to non-linear scales, and we find reliable analytical modeling for the function $R(k)$ up to scales $k \leq 1 $ h/Mpc. In a second modeling context, we modify and propose a new $P(k)$ Halo model, which can fit the simulation data. Finally, we applied our models in a forecast analysis from the perspective of the Euclid survey.  We demonstrate that an Euclid-like survey will possess strong sensitivity to constrain DM-DE coupling, rendering it a crucial probe for this particular class of cosmological models.

The paper is structured as follows: In Section \ref{model}, we delve into the fundamental physical aspects of the IDE model considered in this study. Section \ref{N-body} outlines the methodology and development of our N-body simulation data. Moving forward, in Section \ref{sec:halo_model}, we introduce our adapted Halo model, tailored to accurately fit the N-body simulations. Next, Section \ref{Euclid} presents our forecast analysis, leveraging the sensitivity of an Euclid-like survey. Finally, in Section \ref{sec:conclusions}, we offer our concluding remarks along with future perspectives.

\section{Background expansion and linear perturbations}
\label{model}

The phenomenological possibility of a dark sector interaction has been extensively investigated over the past two decades. Pioneering work in this area was presented in \cite{Amendola_2000, Farrar_2004, Zimdahl_2001, Baldi_2010, Nusser_2005, Amendola_2008}, and it has since been extended to observational tests \cite{Kumar:2016zpg, Murgia:2016ccp, Kumar:2017dnp, DiValentino:2017iww, Yang:2018ubt, Yang:2018euj, Yang:2019uzo, Kumar:2019wfs, Pan:2019gop, Pan:2019jqh, DiValentino:2019ffd, DiValentino:2019jae, DiValentino:2020leo, DiValentino:2020kpf, Gomez-Valent:2020mqn, Yang:2019uog, Lucca:2020zjb, Martinelli:2019dau, Yang:2020uga, Yao:2020hkw, Pan:2020bur, DiValentino:2020vnx, Yao:2020pji, Amirhashchi:2020qep, Yang:2021hxg, Gao:2021xnk, Lucca:2021dxo, deCruzPerez:2023wzd,Kumar:2021eev,Yang:2021oxc,Lucca:2021eqy,Halder:2021jiv,Bernui_2023, Poulin_2023, lucca2021dark, Zhai_2023, Gariazzo_2022, Di_Valentino_2020, mawas2021interacting, Yang_2021, Lucca_2020, Pan_2020, von_Marttens_2020, vanderWesthuizen:2023hcl,Goh:2022gxo,Gomez-Valent:2022bku,Escamilla:2023shf, Forconi:2023hsj, hoerning2023constraints, Wang:2024vmw,Benisty:2024lmj} and N-body simulations \cite{Macci__2004, baldi2010hydrodynamical}. For a comprehensive bibliographic review, see \cite{wang2024further, wang2016dark}. In this section, we will focus on recent developments essential to establishing the theoretical framework central to this work.

Let us start by examining the fundamental characteristics of IDE models. Throughout our discussion, we will operate on the assumption of a spatially flat Friedmann-Lema\^{i}tre-Robertson-Walker (FLRW) metric. 

We can introduce a phenomenological parameterization for the DM-DE interaction within the conservation equations such that the individual stress-energy tensors are no longer conserved, although their sum remains so.\footnote{See~\cite{Nunes:2022bhn} for a more detailed description and~\cite{wang2024further, wang2016dark} for a general review. }

Typically, it is assumed that the covariant derivatives of the DM and DE stress-energy tensors evolve according to:

\begin{eqnarray}
\label{eq:conservDM}
\nabla_{\nu} T^{\mu\nu}_c &=& \frac{Qu^{\mu}_c}{a}\\
\label{eq:conservDE}
\nabla_{\nu} T^{\mu\nu}_x &=& -\frac{Qu^{\mu}_c}{a},
\end{eqnarray}
where $u^{\mu}_c$ denotes the DM 4-velocity vector, and $Q$ represents the DM-DE interaction rate, measured in energy per volume per time.

At this point, we should decide on the functional form of $Q$, and we adopt a widely used choice in the literature, that is $Q = \xi{\cal H}\rho_x$, where ${\cal H}$ signifies the conformal Hubble rate, $\rho_x$ represents the energy density of DE, and $\xi$ serves as a dimensionless parameter governing the strength of the DM-DE interaction. A positive $\xi$ indicates energy transfer from DE to DM, while a negative $\xi$ indicates energy transfer from DM to DE.

When accounting for this specified coupling, the continuity equations dictating the evolution of the energy densities of DM and DE become:
\begin{eqnarray}
\label{eq:continuitydm}
\dot{\rho}_c+3{\cal H}\rho_c &=& \xi{\cal H}\rho_x\, \quad  \\
\label{eq:continuityde}
\dot{\rho}_x+3{\cal H}(1+w_x)\rho_x &=& -\xi{\cal H}\rho_x\, 
\label{eq:continuity}
\end{eqnarray}
where $w_x$ represents the equation of state (EoS) for DE. 
Equations (\ref{eq:continuitydm}) and (\ref{eq:continuityde}) can be analytically integrated, and in the case that both $w_x$ and $\xi$ remain constant throughout cosmic time, they give:

\begin{eqnarray}
\label{eq:solutionrhoc}
\rho_c &=& \frac{\rho_{c,0}}{a^3}+\frac{\rho_{x,0}}{a^3} \left [ \frac{\xi}{3w_x+\xi} \left ( 1-a^{-3w_x-\xi} \right ) \right ]\,; \\
\label{eq:solutionrhox}
\rho_x &=& \frac{\rho_{x,0}}{a^{3(1+w_x)+\xi}} \,,
\end{eqnarray}
where $\rho_{c,0}$ and $\rho_{x,0}$ represent the energy densities of DM and DE today, respectively.

The expression for the Hubble parameter is then given by:

\begin{eqnarray}
\frac{H^2(a)}{H^2_0} = \Omega_{r,0} a^{-4} + \Omega_{b,0} a^{-3} + \Omega_{c,0} a^{-3} \nonumber \\
+ \frac{\Omega_{x,0}}{a^3} \left [ \frac{\xi}{3w_x+\xi} \left ( 1-a^{-3w_x-\xi} \right ) \right ] + \frac{\Omega_{x,0}}{a^{3(1+w_x)+\xi}},
\label{Heq}
\end{eqnarray}
with the first two terms representing the contributions from radiation and baryons, respectively. The relationship between the cosmic and conformal time is defined by ${\cal H} = a H$.

The introduction of the DM-DE coupling further alters the evolution of perturbations. Working in the Newtonian gauge, the coupled system of linear Einstein-Boltzmann equations governing the evolution of dark matter density perturbations, $\delta_c$, and velocity divergences, $\theta_c$, is expressed as follows~\cite{valiviita2008large, honorez2010coupled, gavela2010dark}: 

\begin{eqnarray}
\dot\delta_c &=& -\left( \theta_c - 3\dot\Phi \right) + \frac{Q}{\rho_c}\left(\frac{\delta Q}{Q}  - \delta_c + \Psi \right), \\
\dot\theta_c &=& -{\mathcal{H}}\theta_c + k^2 \Psi.
\label{dm_boltzamnn}
\end{eqnarray}

\begin{figure}[tpb!]
    \centering
    \includegraphics[width=\columnwidth]{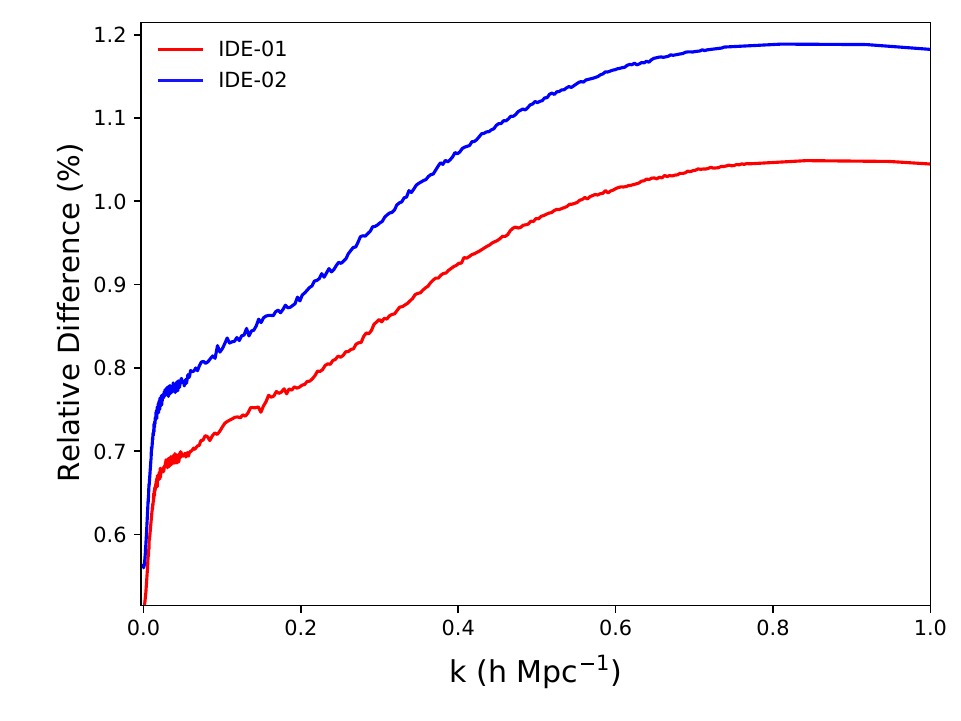} 
    \caption{The relative difference in $P(k, z =0)$, accounting for DE clustering effects, is evaluated within the IDE model discussed in this study.}
    \label{Delta_delta_DE}
\end{figure}

It is worth noting that the Euler equation remains unaltered compared to $\Lambda$CDM, indicating the absence of a fifth force in this model. Additionally, we will not consider the potential presence of clustering DE in this study. As argued in~\cite{Dakin_2019}, DE perturbations may only have discernible effects on very large scales, necessitating a relativistic treatment of simulations. In the regime of Newtonian N-body simulations that we are focused on modeling, DE clustering does not significantly impact the matter power spectrum~\cite{Hassani_2019}. 
Moreover, recent analyses~\cite{dinda2023constraints} have demonstrated that observations favor a homogeneous DE component over a clustering one. 
However, the presence of a homogeneous DE component influences simulations in various ways, including background evolution, growth function calculation, and initial conditions. To provide a more precise quantification within the IDE context, Figure \ref{Delta_delta_DE} illustrates the relative difference in $P(k, z =0)$ with and without DE clustering effects. In constructing the figure, we adopted IDE-01 and IDE-02 baseline values, which will be defined in the subsequent section. It is evident that on large scales, the observed effect is minimal, amounting to less than 1\%. As for scales where non-linear effects are expected to dominate (i.e., $k > 0.2 h/\text{Mpc}$), the relative difference is approximately 1\%. Consequently, we can infer that the perturbations due to DE can be considered negligible. The small oscillations observed in the curves at large scales, often perceived as noise, are attributed to the intrinsic oscillatory nature of $P(k)$ when numerically calculating the ratio between the desired quantities. These fluctuations are negligible from a practical standpoint, as they result in a relative difference of approximately 0.001\% or less. The results were obtained using a customized version of the \texttt{CLASS} code~\cite{blas2011cosmic}, which integrates the IDE model elaborated upon in this study.

On the other hand, within the framework of IDE, the principles of general relativity remain applicable. Consequently, the Poisson equation retains its conventional form:

\begin{equation}
\label{Poisson}
 -k^2 \Psi = 4 \pi G a^2 \rho_c \delta_c.
\end{equation}

It is important to emphasize that in equation (\ref{Poisson}), we have disregarded contributions from DE clustering as argument above. Our focus is currently on sub-horizon scales, where relativistic effects are negligible. Introducing the linear growth function would prove beneficial in this context

\begin{equation}
D(a) = \frac{\delta_c(a)}{\delta_c(1)},
\end{equation}
where $D$ at the present time is unity, $D(1) = 1$.

Following a conventional non-relativistic perturbation-theory approach, which originates from the Boltzmann equation for DM particles (eq. (\ref{dm_boltzamnn})), Poisson's equation (\ref{Poisson}), and the continuity equations for the energy densities of DM (\ref{eq:continuitydm}), we derive the linear growth equation in the presence of interaction between DM and DE species. In this scenario, we obtain:

\begin{equation}
\begin{gathered}
D'' + D' \left[ \frac{\mathcal{H}'}{\mathcal{H}} + \frac{r_{xc}\xi}{a} + 3 \right] \\
- D \left[ \frac{3}{2}\Omega_c - \frac{r_{xc}\xi\mathcal{H}'}{\mathcal{H}} + \frac{r_{xc}\xi}{a^2}(\xi + 3\omega_x + r_{xc}\xi - a)  \right] = 0,
\label{growth_equation}
\end{gathered}
\end{equation}
where we define $r_{\rm xc} = \rho_x/\rho_c$. Here, a prime denotes differentiation concerning the scale factor, denoted as $x' = dx/dln a$. It is noteworthy that when $\xi = 0$, the prediction of the $\Lambda$CDM model is recovered. 

The equation above characterizes the first-order growth factor $D(a)$ and serves as a fundamental equation for N-body gravity simulations. It ensures the accurate implementation of the Zel'dovich approximation at each timestep, utilizing the Zel'dovich equation of motion $x(a) = q + D(a)s$.

\begin{figure}[tpb!]
    \centering
    \includegraphics[width=\columnwidth]{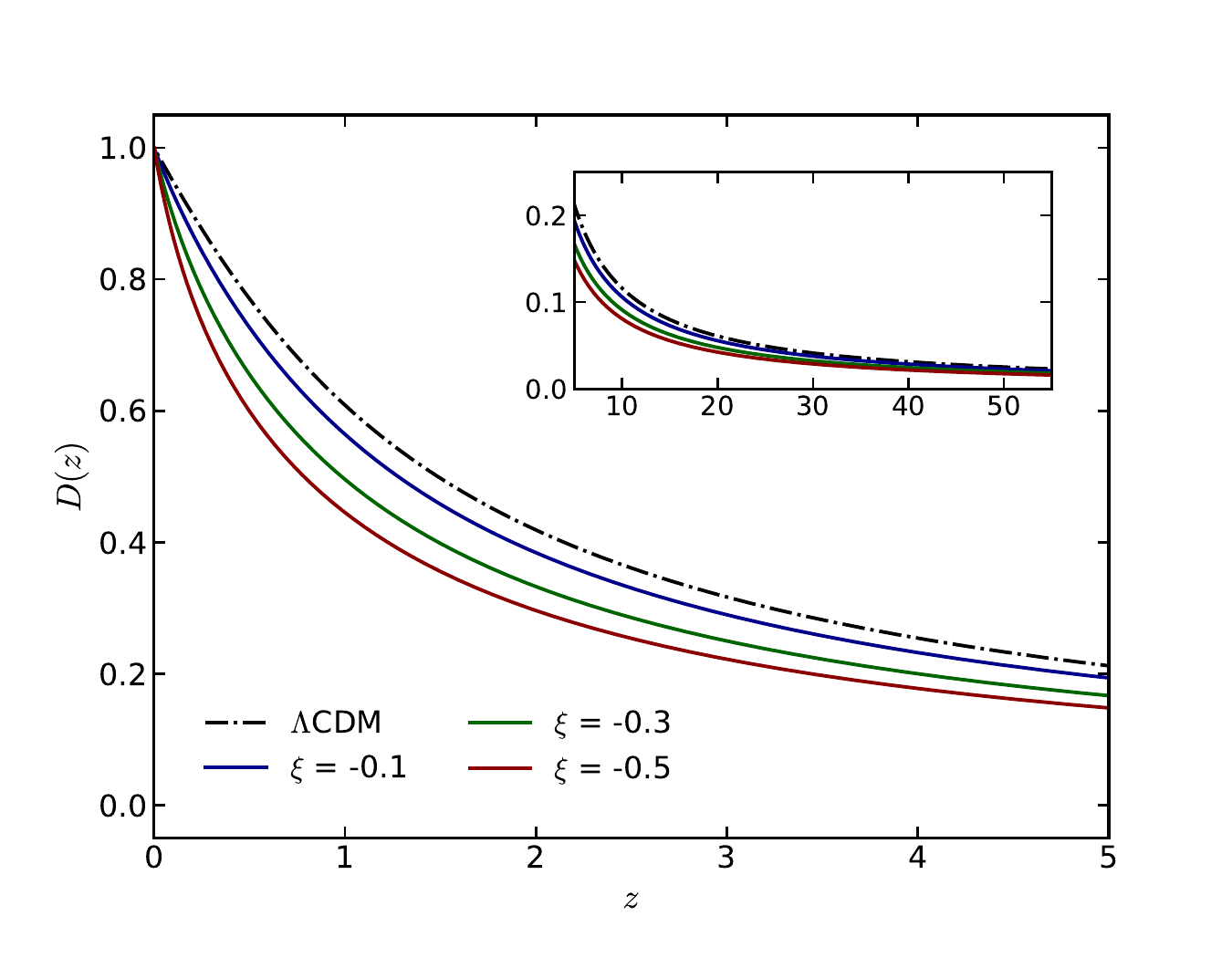} 
    \caption{The growth factor is depicted as a function of redshift. The dashdot black curve represents the standard model, $\Lambda$CDM, whereas the solid curves correspond to the interacting model utilized, showcasing variations of the $\xi$ parameter.}
    \label{D}
\end{figure}

Figure \ref{D} illustrates the growth factor's variation with redshift $z$. It is apparent that as the value of $\xi$ changes, notable differences in the curves' amplitudes arise, particularly at lower redshifts. However, as redshift values increase, the curves gradually converge asymptotically. This trend indicates that within this framework, the interaction between DM and DE significantly influences structures at lower redshifts, whereas at higher $z$ the effects diminish.

\section{N-body simulations}
\label{N-body}

\subsection{Methodology}

The cosmic structure formation in a given theory is most accurately modeled with N-body simulations, where the particles in the simulation are incrementally displaced from their initial positions according to the Poisson equation, the expansion rate of the universe and initial conditions. In this work, we will use the \texttt{pmwd} code~\cite{li2022differentiable}, a differentiable cosmological particle-mesh N-body library. On the other hand, codes based on the Particle-Mesh method, like the \texttt{pmwd} code, may not be as robust as well-established routines, such as those implemented in Gadget \cite{Springel_2005,Springel_2021}, at very small scales. However, when conducting two separate analyses with a finer grid resolution within the same volume, we observed no significant differences in the predicted non-linear power spectrum, $P_{\rm NL}(k)$, within the scales of interest for this work. It is important to note that all our simulations for extracting $P_{\rm NL}(k)$ exclusively involve dark matter particles.

During each simulation, N-body particles discretize the uniform distribution of matter at the beginning
of cosmic history at their Lagrangian positions $q$, from which they then evolve by displacements $s$ to their later positions $\mathbf{x}= q + s(q)$. In what follows, we describe in a nutshell the main physical quantities that are essential, and model-dependent, when running the simulations.

The initial conditions of particles can be set perturbatively when the linear approximation of the density fluctuation is much smaller than 1. We can compute the initial displacements and momenta using the second-order Lagrangian perturbation theory (2LPT)~\cite{bouchet1994perturbative}:

\begin{eqnarray}
    \mathbf{s} = D_1\mathbf{s}^{(1)} + D_2\mathbf{s}^{(2)} ,\nonumber \\
    \mathbf{p} = a^3 H[D_1'\mathbf{s}^{(1)} + D_2'\mathbf{s}^{(2)}],
\end{eqnarray}
where we account for the cosmic background expansion $H$, modified by eq. (\ref{Heq}), allowing the physical position to grow alongside the expansion of the scale factor. 
The amount $\mathbf{p}$ is the canonical momentum for the canonical coordinate $\mathbf{x}$ and $D$ is the growth factor, with the index 1 and 2 representing first (linear) and second order calculations. The input linear growth equation is given for the solution of our eq. (\ref{growth_equation}). 

Following~\cite{li2022differentiable}, we need now to write the growth equation using the suppression factors such that
\begin{align}
    \mathrm{G}_m \equiv D_m/a^m,
\end{align}
where the index $m$ means the order of perturbative expansion in respective quantities. Moreover, the derivatives are given by 
\begin{align}
& D_m=a^m \mathrm{G}_m \\
& D_m^{\prime}=a^m\left(m \mathrm{G}_m + \mathrm{G}_m^{\prime}\right), \\
& D_m^{\prime \prime}=a^m\left(m^2 \mathrm{G}_m+2 m \mathrm{G}_m^{\prime}+\mathrm{G}_m^{\prime \prime}\right) .
\end{align}

The first equation in the 2LPT approximation is then written as

\begin{align}
    &D_1 = a\mathrm{G}_1, \\
    &D_1' = a(\mathrm{G}_1 + \mathrm{G}_1'), \\
    &D_1'' = a(\mathrm{G}_1 + 2\mathrm{G}_1' + \mathrm{G}_1'').
\end{align}
Then we find that the $G$ equation takes the following form for the IDE framework under consideration in this work:

\begin{equation}
\begin{split}
    & \mathrm{G}_1'' + \mathrm{G}_1'\left[ \frac{\mathcal{H}'}{\mathcal{H}} + \frac{r_{xc}\xi}{a} + 5\right] + \mathrm{G}_1 \Bigg[ \frac{\mathcal{H}'}{\mathcal{H}} + \frac{r_{xc}\xi}{a} + 4 \\
    & - \Bigg( \frac{3}{2}\Omega_c - \frac{r_{xc}\xi\mathcal{H}'}{\mathcal{H}} + \frac{r_{xc}\xi}{a^2}(\xi + 3\omega_x + r_{xc}\xi - a)  \Bigg) \Bigg] = 0.
\end{split}
\label{G1}
\end{equation}

The second order follows as

\begin{align}
    &D_2 = a^2\mathrm{G}_2, \\
    &D_2' = a^2(2\mathrm{G}_2 + \mathrm{G}_2'), \\
    &D_2'' = a^2(4\mathrm{G}_2 + 4\mathrm{G}_2' + \mathrm{G}_2''),
\end{align}
and similarly, we obtain

\begin{equation}
\begin{gathered}
\mathrm{G}_2'' + \mathrm{G}_2'\bigg( 4 + \left[ \frac{\mathcal{H}'}{\mathcal{H}} + \frac{r_{xc}\xi}{a} + 3 \right] \bigg) +\\
\mathrm{G}_2\bigg( 4 + 2\Bigg[ \frac{\mathcal{H}'}{\mathcal{H}} +\frac{r_{xc}\xi}{a} + 3 \Bigg] -\\
\Bigg[ \frac{3}{2}\Omega_c  - \frac{r_{xc}\xi\mathcal{H}'}{\mathcal{H}} + \frac{r_{xc}\xi}{a^2}(\xi + 3\omega_x + r_{xc}\xi - a)  \Bigg] \bigg) = \frac{3}{2}\Omega_m \mathrm{G}_1^2,
\end{gathered}
\end{equation}
where the $\mathrm{G}_1$ term is given by eq. (\ref{G1}). This forms a system of coupled equations.

In the cosmological consensus, the overdensity field is a homogeneous and isotropic Gaussian random field characterized by the linear matter power spectrum, $P_{\mathrm{lin}}$,

\begin{align}
\langle \delta^{(1)}(k) \delta^{(1)}(k') \rangle &= (2\pi)^3\delta^D(\mathbf{k}+\mathbf{k'})P_{\mathrm{lin}}(k) \notag\\
&\simeq  \delta^K(\mathbf{k}+\mathbf{k'})VP_{\mathrm{lin}}(k),
\end{align}
where $V$ is a periodic box.

We can calculate the linear contrast $\delta^{(1)}$ by sampling each Fourier mode independently
\begin{equation}
    \delta^{(1)}(k) = \sqrt{VP_{\mathrm{lin}}} \omega(k),
\end{equation}
where $\omega(k)$ is the Fourier transforming of a real white noise field $\omega(q)$.

The input cosmological linear power spectrum can be written as
\begin{equation}
    \frac{k^3}{2\pi^2}P_{\mathrm{lin}}(k) = \frac{4}{25}A_s \left( \frac{k}{k_{\mathrm{pivot}}} \right)^{n_s-1} T^2(k) \frac{c^4 k^4}{\Omega_m^2H_0^4} \left( \frac{D^2(z)}{D^2(0)} \right),
\end{equation}
where $D$ is the linear growth equation, $T$ is the transfer function, $A_s$ is the amplitude of the primordial power spectrum defined at some fixed scale $k_{\mathrm{pivot}}$, and $n_s$ describes the shape of the primordial power spectrum. In summary, one needs to discretize the white noise modes $\omega$, and solve the growth and transfer functions which both depend on a set of cosmological parameters. 

Finally, the core of gravitational N-body simulation is the gravity solver. The gravitational potential sourced by matter density fluctuations follows the Poisson equation, which can be solved efficiently in the Fourier space. As we are not considering modifications in gravitational theory, the Poisson equation, and the gravity solver core remain the same by default. 

To test our framework for the IDE model in N-body simulations introduced above, we shall consider three scenarios: the $\Lambda$CDM model from Planck-CMB best-fit values, and, based on the recent LSS analysis~\cite{Nunes_2022_IDE} in the IDE context, two different input baselines for IDE. Therefore, we will consider the following common fiducial values: $10^9 A_s = 2.1$, $n_s = 0.96$, and $\Omega_b = 0.05$, while we will choose for:

\begin{itemize}
    \item $\Lambda$CDM: $\Omega_m$=0.31, $H_0$=67.36 km/s/Mpc,

    \item IDE-01: $\Omega_m$= 0.29, $H_0$= 68.02 km/s/Mpc, $\xi$ = -0.1,

    \item IDE-02: $\Omega_m$= 0.22, $H_0$= 71.6 km/s/Mpc, $\xi$ = -0.1.
\end{itemize}

For the simulations, we use $256^3$ for the number of particles with single precision and a $512^3$ mesh in a $(256 \,\, {\rm Mpc/h})^3$ box for 63-time steps, from $a = 1/64$ to $a = 1$. These are relatively small simulations but suffice for our purpose in this work. The box size mainly affects the power spectrum at large scales caused by the cosmic variance, and we are interested in modeling $P(k)$ on small scales.  On the other hand, the resolution mainly affects the power spectrum close to the Nyquist limit at small scales. For a better comparison, we apply smoothing with a Gaussian filter to remove the noise in our figures. Hereby a standard deviation of $\sigma = 2.5$ is adopted for the Gaussian kernels in all simulations. For the three scenarios, i.e, $\Lambda$CDM, IDE-01 and IDE-02, we evaluate the non-linear power spectrum, $P(k)$, at $z = 0$. For other values of $z$, the form of $P(k)$ remains unchanged; it merely undergoes a simple amplitude rescaling.

\subsection{Discussion}

To quantify our findings, we introduce the ratio

\begin{eqnarray}
\label{Rnl_model}
R_{\rm NL}(k) = \frac{P_{\rm IDE}(k)}{P_{\rm \Lambda CDM}(k)},
\end{eqnarray}
which quantifies deviations between simulations among the different $P_{\rm NL}(k)$ model predictions.

\begin{figure}[tpb!]
\centering
\includegraphics[width=\columnwidth]{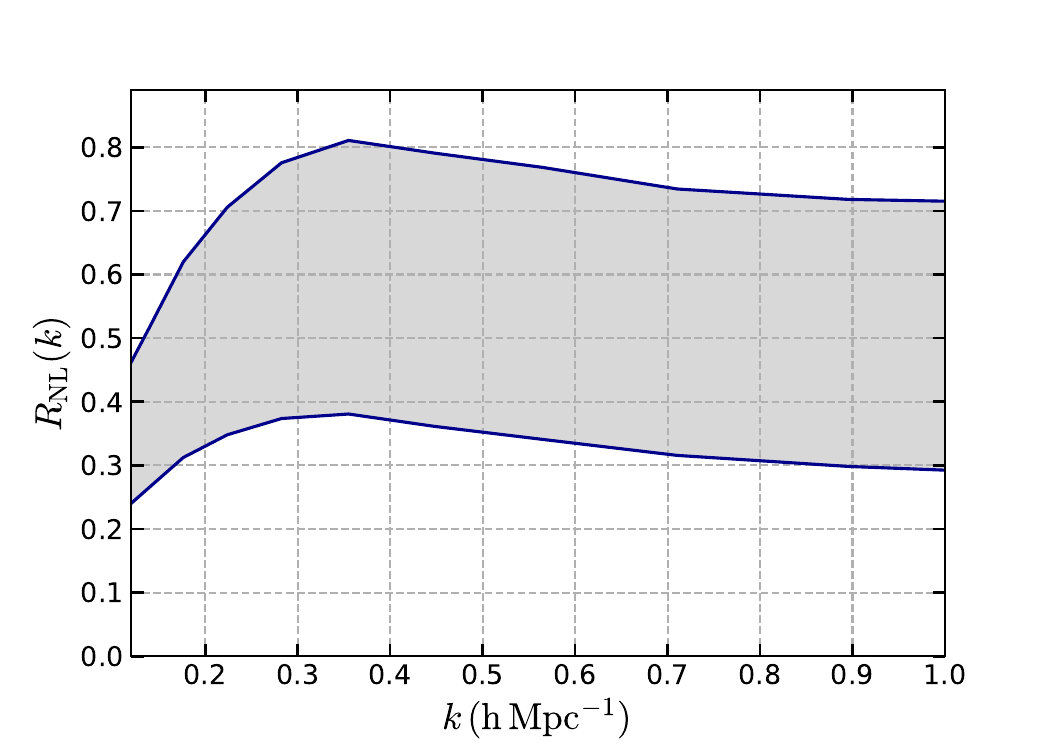}     
\caption{Behavior of the function, eq. (\ref{Rnl_model}), as predicted by N-body simulation data. The upper and lower lines represent deviation limits for IDE-02 (lower line) and IDE-01 (upper line) parameter inputs, respectively.}
\label{Rnl}
\end{figure}

Figure \ref{Rnl} depicts the behavior of the ratio $R_{\rm NL}(k)$, providing insights into the deviation of the IDE model from the $\Lambda$CDM model. The shaded region indicates variations in parameter values between the IDE-01 and IDE-02 baselines. The upper and lower lines highlight the deviation limits predicted for IDE-02 and IDE-01, respectively. As anticipated, IDE-02 demonstrates a more pronounced deviation from $\Lambda$CDM compared to IDE-01. This discrepancy can be attributed to the fact that IDE-01 has only a marginal difference in the chosen parameters with respect to the $\Lambda$CDM case. On the contrary, the important difference between IDE-01 and IDE-02 is not due to a different coupling among DE and DM, but rather a different matter density and Hubble constant.
Between non-linear scales ranging from 0.15 to 0.40 h/Mpc, IDE models exhibit differences of approximately 20\% to 70\% compared to the $\Lambda$CDM model.
Moreover, across these scales, general variations in the IDE baseline parameters may result in deviations of approximately 30\% to 70\% up to scales of about $\sim$ 1 h/Mpc. Thus, it is noteworthy that even minimal variations in the IDE baseline values carry significant implications for predicting $P(k)$ at the simulation level.

Let us also introduce the ratio:

\begin{eqnarray}
R_{\rm IDE}(k) = \frac{P_{\rm NL}(k)}{P_{\rm linear}(k)},
\end{eqnarray}
which quantifies the deviations between the linear and non-linear power spectra as predicted by IDE models themselves.

In our initial approach, we propose to fit the function

\begin{eqnarray}
\label{R_th}
R_{\rm IDE}(k) = A \cdot e^{-n(k - B)^2} + C .
\end{eqnarray}

\begin{figure}[tpb!]
\centering
\includegraphics[width=\columnwidth]{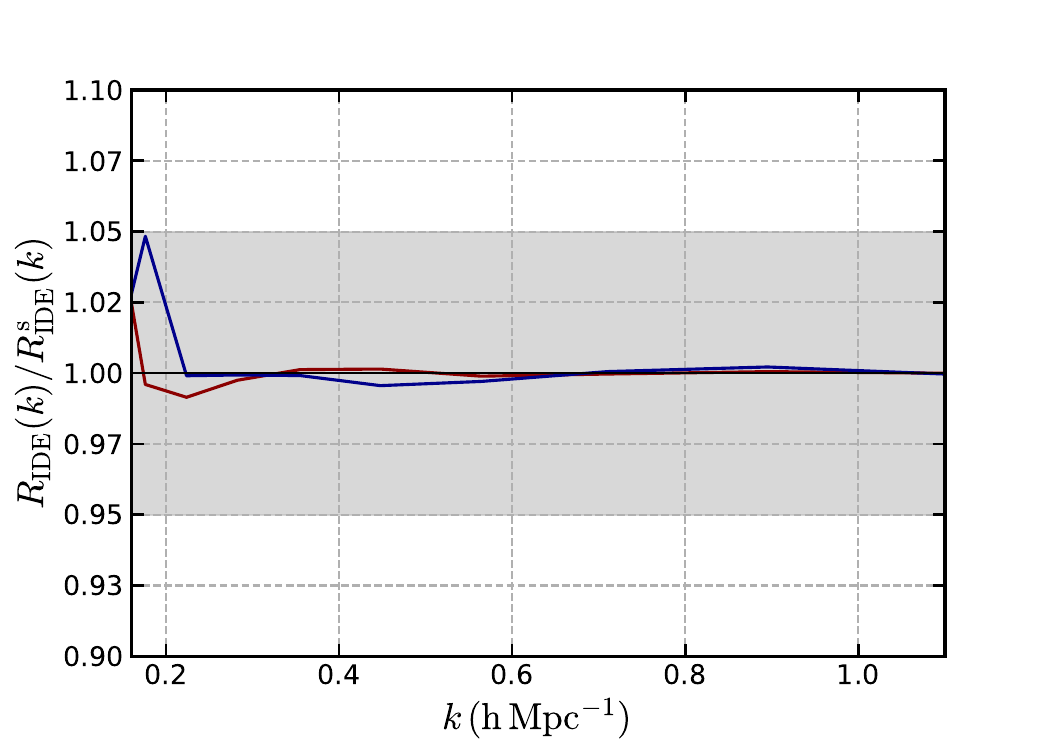}     
\caption{Comparison between predictions of the function with simulation data, \(R_{IDE}^{s}(k)\), and with the fitting function, \(R_{IDE}(k)\). The comparison for IDE-01 (IDE-02) are shown in red (blue), respectively.}
\label{Rk}
\end{figure}

In the following discussion, the superscript $s$ indicates quantities evaluated differently from N-body simulation data. Figure \ref{Rk} illustrates the ratio between $R_{\text{IDE}}^{s}(k)$ and $R_{\text{IDE}}(k)$, parameterized by equation (\ref{R_th}). Our analysis, through chi-square minimization, reveals that the parameters $A \in [23.23, 24.24]$, $B \in [3.77, 4.74]$, $C \in [-0.03,-0.04]$, and $D \in [-9.30, -16.35]$ can effectively fit the non-linear spectra within a $<$5\% relative difference for the values within the range adopted in IDE-01 and IDE-02 samples. It is noteworthy that any observed noise in Figure \ref{Rk} on very large scales is solely attributable to the box size of our simulations. Thus, for the regime where non-linear effects are predominant, i.e., $k > 0.2$ h/Mpc, we note that our fit is very precise with accuracy $<$1\%.

The fitting function, expressed as a parametric approximation like in equation (\ref{R_th}), has been widely employed to fit non-linear scales in phenomenological models. In principle, this approach could suffice to generate valid $P(k)$ outputs within the confines of the simulations conducted. However, in the subsequent section, we will deviate from the conventional methodology and explore a modified Halo model capable of fitting the simulation data. We emphasize that N-body simulations for interacting models, distinct from those explored here, have been extensively investigated in the literature~\cite{baldi2010hydrodynamical, baldi2012codecs, baldi2023cider, hashim2018cosmic, zhang2018fully, palma2023cosmological, zhao2023constraining, liu2022dark, ferlito2022cosmological}.

\section{The Halo Model for IDE}
\label{sec:halo_model}

The leading order moment of the cosmological matter distribution is the non-linear power spectrum, $P_{\rm NL}(k)$. The development and results presented in the previous section offer a parametric function that relates the predictions $P_{\rm NL}(k)$ and $P_{\rm linear}$. While this approach is commonly used in the literature, we also focus on a second argument: the development of a new halo model that can incorporate the model of interest in this work.

To do this, we used the structure based on the \texttt{HMCode}~\cite{Mead_2016} (see~\cite{asgari2023halo} for a review of halo models). By incorporating simulation-calibrated quantities, such as the halo mass function and dark density profiles adjusted for baryonic feedback, the halo model offers a robust integration between theoretical predictions and observational data. This approach is especially effective in modeling the nonlinear power spectrum, $P_{\rm NL}(k)$, and allows the model to be reused for other analyses without the need for additional simulations, which are expensive in terms of cost and computation.

In the halo model, the matter power spectrum $P(k)$ is the contribution of two parts. The first part is associated with the description of non-linearities and is called the one-halo term, $P_{\rm 1H}$. The second part is called the two-halo term and describes the linear part of the spectrum, associated with small values of $k$, $P_{\rm 2H}$. Thus, we have that the total power spectrum is given by

\begin{eqnarray}
P(k) = P_{\rm 1H} + P_{\rm 2H}.
\end{eqnarray}

In many cases, it is also convenient to use dimensionless quantities $\Delta^2(k)=4\pi V\left(\frac k{2\pi}\right)^3P(k)$. For the first term, it is observed that the power spectrum presents the characteristics of a shot noise, moderated by the halo density profile

\begin{eqnarray}
\Delta_{1\text{H}} ^ 2 ( k ) = \left ( \frac k { 2 \pi }\right)^3\frac{4 \pi}{\overline{\rho}^2}\int_0^\infty M^2W^2(k,M)F(M)\mathrm{d}M .
\label{term-1}
\end{eqnarray}

In the above equation, the power spectrum is formulated as an integral over halo masses, $M$, with the differential halo mass function denoted as \(dM\), determined by the halo mass function, \(F(M)\), which predicts the abundance of dark matter halos of different masses. Additionally, this integral incorporates the term \(W(k, M)\), representing the normalized Fourier transform of the halo density profile~\cite{asgari2023halo}. On larger spatial scales, the distribution of haloes exhibits non-random patterns, and the displacements between them necessitate the incorporation of a two-halo term into the power spectrum analysis. This term accounts for the interactions between haloes and is akin to the linear-theory power spectrum for the matter distribution, $\Delta_{2\mathrm{H}}^{2}(k)=\Delta_{\mathrm{lin}}^{2}(k)$.

The \texttt{HMCode} exhibits specific modifications about the halo model, involving parameters fine-tuned through simulations and functions optimized to align with them. Concerning the power spectrum, an adaptation is introduced to the one-halo and two-halo terms. The one-halo term is enhanced with a function to more effectively tailor the spectrum at these scales. Consequently, it is redefined as follows:

\begin{eqnarray}
\bar{\Delta}_{2\mathrm{H}}^{2}(k)=\left[1-f\tanh^{2}(k\sigma_{\mathrm{V}}/\sqrt{f})\right]\Delta_{\mathrm{lin}}^{2}(k),
\end{eqnarray}
where \(f\) represents a free parameter, \(\sigma_{\mathrm{V}}\) is the 1D linear-theory displacement variance, and \( \Delta_{\mathrm{lin}}^2(k) \) refers to the adaptation of the linear term made by~\cite{crocce2006transients} in the form

\begin{eqnarray}
\Delta_{\mathrm{lin}}^2(k)\to\mathrm{e}^{-k^2\sigma_\mathrm{v}^2}\Delta_{\mathrm{lin}}^2(k).
\end{eqnarray}

The one-halo term has also undergone modifications; in this instance, it exhibits a more rapid decay and is consequently redefined as:

\begin{eqnarray}
\bar{\Delta}_{1\text{H}}^{2}=[1-\mathrm{e}^{-(k/k_{*})^2}]\Delta_{1\text{H}} ^ 2,
\end{eqnarray}
where $k_{*}$ represents a free parameter associated with the one-halo damping wavenumber. The complete spectrum incorporates an additional $\alpha$ term, markedly enhancing the transition between the respective terms. Its revised definition is as follows:

\begin{eqnarray}
\Delta^{2}(k)=[(\bar{\Delta}_{2\text{H}}^{2})^{\alpha}+(\bar{\Delta}_{1\text{H}}^{2})^{\alpha}]^{1/\alpha}.
\end{eqnarray}

Additionally, the halo model incorporates key elements crucial to its formulation. One of these elements is the halo mass density profile $\rho(r, M)$, which is selected using the Navarro-Frenk-White (NFW) profile~\cite{navarro1997universal}. 

Moreover, another significant component is the halo mass function (HMF), denoted as $F(M)$, which can be universally represented by $f(\nu)$:

\begin{equation}
\frac{M}{\overline{\rho}}F(M)\mathrm{d}M=f(\nu)\mathrm{d}\nu. 
\end{equation}

Here, $\nu$ represents the peak threshold of a halo with mass $M$, defined by:

\begin{equation}
 \nu = \frac{\delta_c(z)}{\sigma_8(M,z)}, 
\end{equation}
where $\delta_c(z)$ denotes the critical threshold for spherical collapse based on linear theory, and the variance $\sigma_8(M,z)$ is computed from the linear matter power spectrum. In the \texttt{HMCode}, the improved formula developed by~\cite{sheth1999large} is applied to determine \( f(\nu) \). 

Now, it is crucial to understand the model dependency up to this point. By default, any non-standard cosmology will automatically change the predictions on $\sigma_{8}(M,z)$ and $\Delta_{\mathrm{lin}}^{2}(k)$. On the other hand, $\delta_{c}(z)$ is also a quantity depending on the model. Therefore, we changed the default prediction to $\delta_{c{\rm IDE}} = ({\texttt{HMCode}})^{\beta_1}$, where $\texttt{HMCode}$ represents the default term implemented in the code. The changes in $\sigma_{8}(M,z)$ and $\Delta_{\mathrm{lin}}^{2}(k)$ are already automatically captured by the modifications described in Section~\ref{model}.

Another essential quantity in the formulation of halo models, which depends on the adopted cosmology, is the concentration-mass relationship $c(M,z)$. In this context, Ref.~\cite{dolag2004numerical} underscores the necessity of incorporating a correction factor into the value. This correction factor involves multiplying the value by the ratio of the linear growth function associated with the chosen cosmology to that of the standard model. This adjustment accounts for exotic cosmological scenarios. For the IDE model, our proposition involves:

\begin{equation}
c(M,z)= A \frac{1+z_\mathrm{f}(M)}{1+z} \times \frac{g_{\rm IDE}(z\to\infty)}{g_\mathrm{\Lambda CDM}(z\to\infty)},
\label{concentration}
\end{equation}
where $z_f$ represents the redshift at which halo formation occurs, a quantity that depends on the mass and is further detailed in~\cite{bullock2001}. Additionally, $g_{\rm IDE}$ depends on the growth index, which is described by 

\begin{equation}
\frac{\mathrm{d}\ln g}{\mathrm{d}\ln a}=\Omega_{\mathrm{m}}^{\gamma_{\rm IDE}}(z).
\end{equation}

Notably, the growth index we employ deviates from standard values, as elucidated in~\cite{marcondes2016analytic}:

\begin{equation}
\gamma_{\rm IDE} = \frac{3(1-\omega_0 + 5 \xi)}{5-6\omega_0-6\xi}.
\end{equation}

We have also made a crucial modification in the term $(\bar{\Delta}_{2\text{H}}^{2})$, redefining it for the IDE model as:

\begin{equation}
\bar{\bar{\Delta}}_{2\mathrm{H}}^{2}(k) = \beta_2 \bar{\Delta}_{2\mathrm{H}}^{2} \mathrm{e}^{k \beta_3},
\end{equation}
where \( \beta_2  \) and \( \beta_3 \) are free parameters to be calibrated by simulation data. 

Another aspect that we note in the IDE framework is the location and magnitude of the transition that occurs in the one-halo and two-halo terms. To mitigate these impacts, the value of $k_{*}$ at $\bar{\Delta}_{1\text{H}}^{2}$ is modified to the new ones in the form $k_{* \rm IDE} = ({\texttt{HMCode}}){\beta_4}$, and the value of $\alpha$ is also changed to $\alpha_{\rm IDE} = ({\texttt{HMCode}}){\beta_5}$.

Taking these modifications into account, our complete matter power spectrum is given by

\begin{eqnarray}
\Delta^{2}(k)=[(\bar{\bar{\Delta}}_{2\text{H}}^{2})^{\alpha_{\rm IDE}}+(\bar{\Delta}_{1\text{H}}^{2})^{\alpha_{\rm IDE}}]^{1/\alpha_{\rm IDE}}.
\end{eqnarray}

In summary, our adaptation of the Halo Model for IDE introduces five new fitting parameters, namely $\beta_1$, $\beta_2$, $\beta_3$, $\beta_4$, and $\beta_5$. These parameters essentially quantify deviations that are contingent upon cosmological models within the framework of the halo model and need to be determined by N-body simulation data.

\begin{figure}[tpb!] 
\centering
\includegraphics[width=\columnwidth]{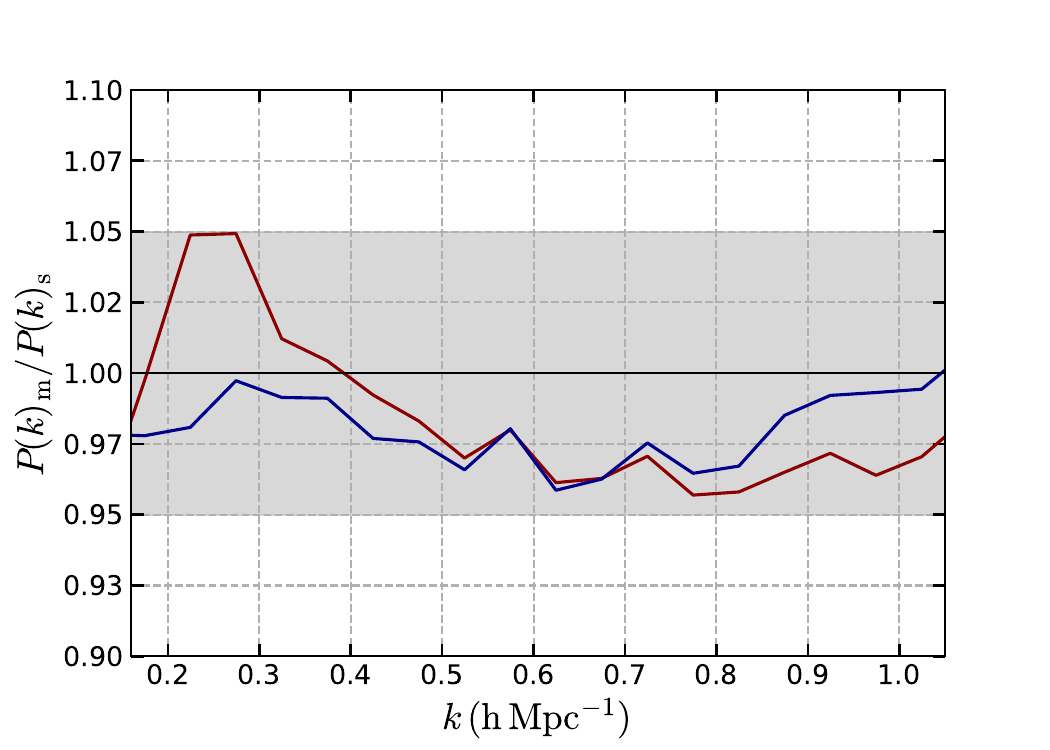}     
\caption{Ratio between the non-linear matter power spectrum generated by N-body simulation, \(P(k)_s\), and our adapted model \(P(k)_m\). The comparison for IDE-01 is shown in red and for IDE-02 in blue.}
\label{pk}
\end{figure}

Our analysis indicates that $\beta_1 \in [3.4, 3.6]$, $\beta_2 \in [0.15, 0.28]$, $\beta_3 \in [0.15, 1.31]$, $\beta_4 \in [ 3.5, 5.1]$ and $\beta_5 \in [1.09, 1.11]$ can fit with good accuracy the entire parameter-space range between the IDE-01 and IDE-02 model baselines. The determination of the fitting values involved an extensive study of the \texttt{HMCode} already implemented in \texttt{CLASS}, where parameter variations were analyzed to assess the impact of the IDE cosmology on the key functions of the adopted halo model. The creation of these parameters played a crucial role in our model's ability to accurately describe the power spectra of matter on non-linear scales.
Figure \ref{pk} illustrates the ratio between the modified halo model and simulation data for two scenarios (IDE-01 and IDE-02). Notably, modeling accuracy within a 5\% margin extends to values of \(k \approx 1\). A relative difference of less than 5\% compared to simulations is a promising outcome, suggesting that our model may be robust enough to capture the subtleties of matter structures even in different settings, as tested with IDE-01 and IDE-02. It is important to note that the Euclid survey aims to measure power spectra with 1\% accuracy at linear scales. Our current calibration achieves an adjustment within 5\% for scales up to approximately 1 h/Mpc. Given the complexity of the phenomenological model under consideration in this work, and considering that our proposal is pioneering—specifically, in developing a new halo model for the IDE scenario—we plan to offer further refinements in a future publication, incorporating a significantly larger number of simulations and enhanced precision. 

In the following sections, we implement our halo model in a forecast analysis based on the sensitivity of the Euclid survey.

\section{Forecast analysis from the Euclid Survey perspective}
\label{Euclid}

Building upon the development of the matter power spectrum model that captures the effects of dark sector interactions on non-linear scales, as outlined in the previous section, we now turn our attention to a forecast analysis. In this section, we aim to leverage the sensitivity of galaxy clustering and cosmic shear measurements from the Euclid mission. This analysis will primarily focus on quantifying the accuracy and validation of our developed model. Our forecast analysis methodology is based on the approach presented in~\cite{sprenger2019cosmology}.

We know that the arrangement of galaxies in space provides a good tracer of the hidden distribution of dark matter. The relationship between the power spectrum of galaxies, $P_g$, and the power spectrum of matter, $P_m$, is given by:

\begin{equation}
\begin{gathered}
P_g(k,\mu,z)=f_\text{AP}(z)\times f_\text{res}(k,\mu,z)\times\\ f_\text{RSD}(k,\mu,z)\times b^2(z)\times P_m(k,z),
\label{pg}
\end{gathered}
\end{equation}
where $f_\text{AP}(z)$ is the term associated with the Alcock-Paczynski effect, $f_\text{res}(k,\mu,z)$ is a suppression factor, $b(z)$ is the bias function (for more details see~\cite{sprenger2019cosmology}) and $f_\text{RSD}$ incapsule the effects of the redshift space distortions. The inclusion of the last term mentioned is crucial for the accuracy of the analysis.

It turns out that the redshift information we receive is not solely generated by the expansion of the universe. The Doppler effect, arising from the peculiar random velocities of galaxies, also contributes, creating a distorted appearance in the redshift pattern known as the `fingers of God'~\cite{jackson1972critique}. Here, we describe these distortions using an exponential suppression factor, as outlined in~\cite{bull2015late}, so that the third term of equation \ref{pg} is given by

\begin{equation}
f_{\mathrm{RSD}}(k,\mu,z)=\left(1+\eta(k,z)\mu^2\right)^2e^{-k^2\mu^2\sigma_{\mathrm{NL}}^2},
\end{equation}
where the first term in parentheses represents the Kaiser formula~\cite{kaiser1987clustering}, while the exponential term addresses the `fingers of God' phenomenon. Specifically, $\sigma_{\mathrm{NL}}$ is initially set at 7 Mpc, with a range of 4-10 Mpc considered in our predictions. Meanwhile, $\eta$ represents the corrected growth rate, potentially influenced by the galaxy bias:

\begin{equation}
\eta(k,z)=-\frac{1+z}{2b(z)}\cdot\frac{\mathrm{d}\ln P_m(k,z)}{\mathrm{d}z}.
\end{equation}

Continuing with the discussion of eq. (\ref{pg}), while it provides a comprehensive definition by incorporating various factors to ensure accurate measurements, it is crucial to account for the impact of experimental noise. As a result, we partition our surveys into bins of width $\Delta z = 0.1$, centered around the mean redshift $\bar{z}$. Within each bin, we compute correlation functions to investigate the power spectrum at a specific fixed redshift $\bar{z}$. The volume of each redshift bin is determined as follows:

\begin{equation}
V_r(\bar{z})=\frac{4\pi}{3}f_{\mathrm{sky}}\cdot\left[r^3\left(\bar{z}+\frac{\Delta z}{2}\right)-r^3\left(\bar{z}-\frac{\Delta z}{2}\right)\right],
\label{volume}
\end{equation}
where $f_{\text{sky}}$ represents the fraction of the survey's coverage over the sky. Subsequently, to account for this experimental noise, the observed quantity in each bin is adjusted as:

\begin{equation}
P_\text{obs}(k,\mu,\bar{z})=P_g(k,\mu,\bar{z})+P_N(\bar{z}),
\end{equation}
where $P_N(\bar{z})$ denotes the volume expressed in eq. (\ref{volume}) divided by the number of galaxies in the respective bin.

We will focus on redshifts ranging from $0.7$ to approximately $2$, with a sky coverage of 0.3636. In this analysis, the spectroscopic redshift measurement accuracy, $\sigma_z$, is impressively low at $0.001(1+z)$. This small error is crucial because it translates into a minimal radial distance uncertainty, $\sigma_{||}$, via the formula $\sigma_{||} = c \sigma_z / H$.

Furthermore, the bias factor for Euclid-detected galaxies is approximated by a simple linear relation, $b(z)=1+z$. To address potential deviations from this relation, let us also consider two nuisance parameters, each with a mean value of 1:

\begin{equation}
b(z)=b_0 (1 + z)^{0.5 b_1},
\end{equation}
where in our analysis we will impose a stringent 5\% precision, 2$\sigma$ CL, constraint on these $b$-factors to ensure accurate modeling.

In our forecast analysis, we will incorporate the cosmic shear survey, which captures the alignments of galaxies induced by weak gravitational lensing resulting from large-scale structures along the line of sight. We will evaluate this phenomenon in terms of the angular power spectrum, as detailed in~\cite{sprenger2019cosmology} for the Euclid survey. In this step, we integrate our three-dimensional matter power spectrum model into these calculations to enhance the accuracy of our predictions. To define the $\chi^2$, we adopt an expression akin to that presented in~\cite{hilbert2017intrinsic}, given by 

\begin{equation}
\begin{gathered}
\chi^2=\sum_n\int_{k_{\min}}^{k_{\max}}\mathrm{d}k\cdot k^2\int_{-1}^1\frac{V_r(\bar{z}_n)}{2(2\pi)^2}\\
\times \left[\frac{(\Delta P_g(k,\mu,\bar{z}_n))^2}{\sigma_{\mathrm{obs}}^2(k,\mu,\bar{z}_n)+\sigma_{\mathrm{th}}^2(k,\mu,\bar{z}_n)}\right]\mathrm{d}\mu,
\label{chi2}
\end{gathered}
\end{equation}
where $\Delta P_g(k,\mu,\bar{z}_n)$ denotes the difference between the predicted and observed galaxy power spectrum and to bolster the analytical rigor, we also take into account the term denoted as $\sigma_{th}$ (of which take the role of correlation lengths $\Delta k$, $\Delta \mu$, $\Delta z$), given by

\begin{equation}
\sigma_\mathrm{th}(k,\mu,z)=\left[\frac{V_r(z)}{(2\pi)^2}k^2\Delta k\frac{\Delta z}{\Delta\bar{z}}\right]^{1/2}\alpha(k,\mu,z)P_g(k,\mu,z).
\end{equation}

For the correlation length in wavenumber space, we set $\Delta k = 0.05 h/\text{Mpc}$. This choice aligns closely with the BAO scale, which represents the smallest inherent scale in the matter power spectrum, thus providing a conservative estimate for the correlation length in k-space. The purpose of the relative error envelope function $\alpha(k, \mu, z)$ is to primarily address uncertainties in two types of non-linear corrections: the prediction of the matter power spectrum itself and the bias (see discussion in~\cite{sprenger2019cosmology}).


\begin{table}[h]
\centering
\begin{tabular}{cccc} 
\toprule
\textbf{Parameter} & \multicolumn{3}{c}{$\mathbf{\sigma (\text{Euclid})}$} \\
\cmidrule(lr){2-4}
 & $\mathbf{k_{max} = 0.1}$ & $\mathbf{k_{max} = 0.5}$ & $\mathbf{k_{max} = 1}$ \\ 
\midrule
$\Omega_{\rm cdm}$ & 0.037 & 0.018 & 0.015 \\
$n_s$ & 0.019 & 0.0065 & 0.0055 \\
$h$ & 0.0169 & 0.0085 & 0.0066 \\
$\xi$ & 0.0136 & 0.0131 & 0.0110 \\
$\sigma_{\rm NL}$ & 0.137 & 0.072 & 0.059 \\
$b_0$ & 0.0108 & 0.0039 & 0.0032 \\
$b_1$ & 0.0143 & 0.0066 & 0.0058 \\
\bottomrule
\end{tabular}
\caption{Summary of the expected 1$\sigma$ sensitivity of Euclid to the cosmological parameters. The parameters $b_0$ and $b_1$ represent the galaxy bias parameters, and $\sigma_{\rm NL}$ accounts for the effects of RSD. The parameter $h$ is the reduced Hubble parameter, i.e., $h = H_0/100$.}
\label{tab:example}
\end{table}

\begin{figure*}
         \centering
        \includegraphics[width=\textwidth]{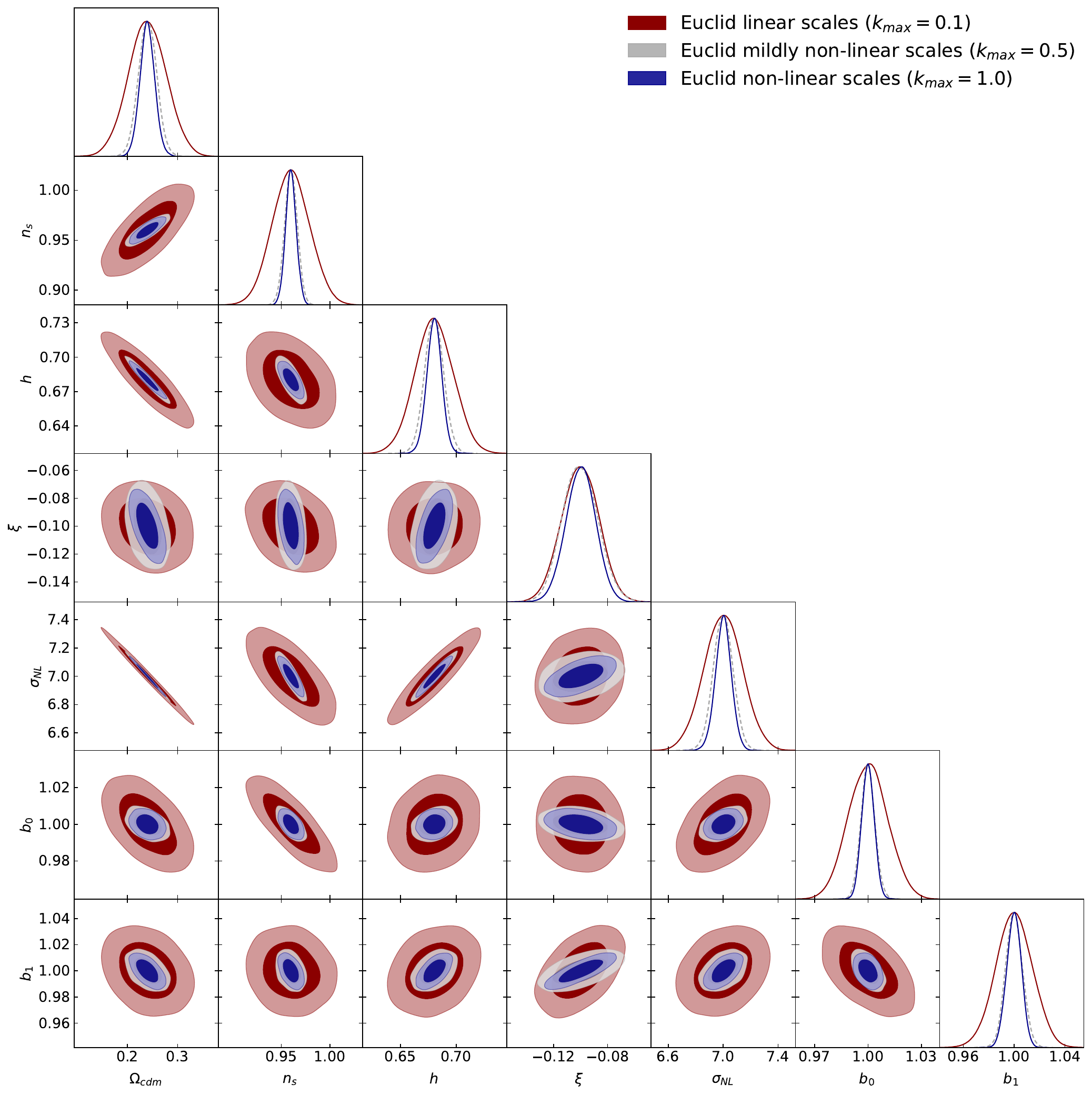}
        \caption{Marginalized 1$\sigma$ and 2$\sigma$ contours and one-dimensional posteriors in the parameter space, showing the expected sensitivity of Euclid to the cosmological parameters for the IDE-01 scenario baseline values \textbf{for different cutting scales $k{\max}$}. The analysis is performed following the conservative approach for the description of the theoretical error.}
         \label{fig:PS_euclid}
\end{figure*}

In what follows in our main results, we use a modified version of the \texttt{CLASS}$+$\texttt{MontePython} code~\cite{blas2011cosmic, audren2013conservative, brinckmann2019montepython}, adapting the likelihoods \textit{euclid\_pk} as well as incorporating all background dynamics, linear perturbations, and the modified halo models in \texttt{CLASS}, as discussed in the previous sections. 

In our forecast analyses, we considered three cases by applying different cut-off scales: the linear regime ($k_{\rm max} = 0.1 h/\text{Mpc}$), the mildly non-linear ($k_{\rm max} = 0.5 h/\text{Mpc}$), and the non-linear ($k_{\rm max} = 1 h/\text{Mpc}$). Our model exhibits good predictive capability across these scales, successfully fitting the simulations. Furthermore, we incorporate baryon feedback using a six-parameter physical model, which includes AGN-driven gas ejection and star formation processes, as implemented in \texttt{HMCode}~\cite{Mead:2020vgs}. This model, validated against hydrodynamic simulations, accurately captures feedback effects in the nonlinear power spectrum with sub-percent precision, even on very small scales. For a recent discussion on the significance of baryonic effects in the power spectrum, see also \cite{Parimbelli_2019, Chisari_2019, terasawa2024exploringbaryoniceffectsignature, xu2023constrainingbaryonicphysicsdes, Salcido_2023}.

Table \ref{tab:example} presents a statistical summary detailing the anticipated sensitivity of the Euclid mission to cosmological parameters at $1\sigma$ confidence level (CL). These results provide invaluable insights into the precision with which Euclid can probe the IDE framework adopted in this work. Throughout our analysis, we have adopted the IDE-01 scenario baseline values, ensuring consistency and comparability across our findings. This choice serves as a robust foundation for our discussions, enabling us to explore the nuanced implications of Euclid's observations. Importantly, our sensitivity constraints remain robust and independent of specific input values for the parameters baseline. This inherent flexibility underscores the generality of our results, allowing for broad applicability and facilitating meaningful interpretations across this IDE framework.

Of particular significance is the examination of sensitivity towards the coupling parameter $\xi$. According to the analysis of Euclid forecasts, the coupling parameter can be constrained with a precision of $\sigma(\xi) = 0.013$ (0.011) considering only linear (non-linear) scales, respectively. This represents a substantial improvement by an order of magnitude compared to any other current observational tests documented in the literature. For example, in the study by~\cite{Nunes_2022_IDE}, which employed the full-form galaxy power spectrum of the BOSS DR12 sample in conjunction with CMB data from Planck, a constraint of $\xi > -0.12$ was established. Consequently, the forthcoming data from Euclid has the potential to enhance these constraints by up to two orders of magnitude. Therefore, the accuracy of the constraints generated by the Euclid mission sensitivity will be crucial for the IDE models. The joint accuracy with $\sigma(\Omega_{\rm cdm})$ and $\sigma(h)$, two parameters highly sensitive to IDE dynamics, will also be constrained with remarkable precision. Consequently, Euclid data alone will possess the capability to constrain the dynamics of IDE models, encompassing both background and perturbative levels, with unprecedented accuracy.

Figure \ref{fig:PS_euclid} show the marginalized 1$\sigma$ and 2$\sigma$ contours and one dimensional posteriors in the parametric space: $\Omega_{\rm cdm}$, $n_s$, $h$, $\xi$, $\sigma_{\rm NL}$, $b_0$, $b_1$. Here, $h$ represents the reduced Hubble parameter, defined as $h = H_0/100$. In the following discussions, we will adhere to this notation for the Hubble constant.
Let us refocus our attention on the coupling parameter. As previously noted, the anticipated high-sensitivity forecast, which can achieve an accuracy of $\sigma(\xi) = 0.0110$, suggests that with the current observed value of $\xi = -0.1$, there could potentially be an indication of interaction within the dark sector. To generalize and repeat the analysis with a different input value of $\xi=0$, we find $\xi > -0.01$ at 1$\sigma$ CL. Therefore, it will be crucial to integrate forthcoming real data from Euclid with existing data from the literature for a real comprehensive analysis. 

There are other noteworthy points to consider. In forecast analyses from the Euclid perspective, a persistent and significant negative correlation between the coupling parameter $\xi$ and the Hubble constant $h$ is observed. This correlation has been pivotal in supporting the argument that this class of models might resolve the $H_0$ tension~\cite{Zhai_2023, Bernui_2023, Gariazzo_2022}. Interestingly, the analysis also suggests that $\xi$ may exhibit a notable new correlation with the galaxy bias parameters $b_0$ and $b_1$ (see Figure \ref{fig:PS_euclid}). Moreover, the observed correlation in the $\xi$-$h$ plane within the IDE framework implies that the galaxy bias parameters are likely to correlate with $h$ as well. This insight unveils a previously unrecognized cosmological characteristic of these models, which will be crucial to understand in future analyses involving real observational data.

\section{Final Remarks}
\label{sec:conclusions}

In conclusion, our study represents a significant step forward in the utilization of IDE models for observational constraints within the realm of LSS data analysis on non-linear scales. By leveraging N-body simulations and refining a modified halo model tailored to IDE cosmologies, we have demonstrated the potential for achieving remarkable precision in fitting simulated spectra down to scales of approximately $\sim$ 1 h/Mpc. These findings not only enhance our understanding of the dynamics of dark energy interactions but also underscore the viability of IDE models as powerful tools for cosmological analysis.
Moreover, our work opens new avenues for future research, with implications for upcoming observational surveys such as the Euclid mission. The robustness of our model, as demonstrated through forecast analyses, suggests its potential utility in informing and refining cosmological parameter estimates from forthcoming observational data.

In anticipation of future developments, our next steps involve the application and refinement of our models on real observational data, including datasets such as the Dark Energy Survey~\cite{Abbott_2022, survey2023y3+} and the Kilo-Degree Survey (KIDS-1000)~\cite{Asgari:2020wuj}. Specifically, we aim to investigate the robustness of our model in addressing the current tension in the $S_8$ parameter. We are actively working on analyzing the results in this regard, and we look forward to sharing our findings in future communications.

\begin{acknowledgments}
\noindent The authors express their gratitude to the referee for their valuable comments and suggestions, which have significantly improved the quality of this work. We thank Carsten van de Bruck and Yuejia Zhai for useful discussions. E. S. thanks CAPES for partial financial support. R.C.N. thanks the financial support from the Conselho Nacional de Desenvolvimento Cient\'{i}fico e Tecnologico (CNPq, National Council for Scientific and Technological Development) under the project No. 304306/2022-3, and the Fundação de Amparo à Pesquisa do Estado do RS (FAPERGS, Research Support Foundation of the State of RS) for partial financial support under the project No. 23/2551-0000848-3.  E.D.V. is supported by a Royal Society Dorothy Hodgkin Research Fellowship. This article is based upon work from COST Action CA21136 Addressing observational tensions in cosmology with systematics and fundamental physics (CosmoVerse) supported by COST (European Cooperation in Science and Technology).

\end{acknowledgments}

\bibliography{PRD}

\end{document}